\documentclass{jfm}

\usepackage{graphicx}
\usepackage{natbib}

\ifCUPmtlplainloaded \else
  \checkfont{eurm10}
  \iffontfound
    \IfFileExists{upmath.sty}
      {\typeout{^^JFound AMS Euler Roman fonts on the system,
                   using the 'upmath' package.^^J}%
       \usepackage{upmath}}
      {\typeout{^^JFound AMS Euler Roman fonts on the system, but you
                   dont seem to have the}%
       \typeout{'upmath' package installed. JFM.cls can take advantage
                 of these fonts,^^Jif you use 'upmath' package.^^J}%

      }
  \else
  \fi
\fi

\ifCUPmtlplainloaded \else
  \checkfont{msam10}
  \iffontfound
    \IfFileExists{amssymb.sty}
      {\typeout{^^JFound AMS Symbol fonts on the system, using the
                'amssymb' package.^^J}%
       \usepackage{amssymb}%

      }{}
  \fi
\fi

\ifCUPmtlplainloaded \else
  \IfFileExists{amsbsy.sty}
    {\typeout{^^JFound the 'amsbsy' package on the system, using it.^^J}%
     \usepackage{amsbsy}}
    {}
\fi

\def\beq{\begin{equation}}
\def\eeq{\end{equation}}

\def\beq{\begin{equation}}                           
\def\eeq{\end{equation}}                           
\def\bea{\begin{eqnarray}}                           
\def\eea{\end{eqnarray}}        

\def\rp{\mathbf{r}_{\perp}}

\newsavebox{\astrutbox}
\sbox{\astrutbox}{\rule[-5pt]{0pt}{20pt}}

\title[A Drop of Active Matter]{A Drop of Active Matter}

\author[J.-F. Joanny and S. Ramaswamy]%
{JEAN-FRAN\c{C}OIS JOANNY$^1$%
  \thanks{E-mail: {\tt jean-francois.joanny@curie.fr}}\ns
\and SRIRAM RAMASWAMY$^2$
\thanks{E-mail: {\tt sriram@physics.iisc.ernet.in}; also at JNCASR, Bangalore 560 064, India}
}
\affiliation{$^1$Physicochimie Curie (CNRS-UMR168, UPMC Universit\'{e} Paris VI), Institut Curie, Section de Recherche
26 rue d'Ulm, 75248 Paris Cedex 05, France\\[\affilskip]
$^2$Centre for Condensed Matter Theory, Department of Physics, Indian Institute of Science, Bangalore 560 012, India} 
\pubyear{2010}
\volume{650}
\pagerange{119--126}
\date{?; revised ?; accepted ?. - To be entered by editorial office}
\begin{document}

\maketitle

\begin{abstract}
We study theoretically the hydrodynamics of a fluid drop containing oriented filaments endowed with active contractile or extensile stresses and placed on a solid surface. The active stresses alter qualitatively the wetting properties of the drop, leading to new spreading laws and novel static drop shapes. Candidate systems for testing our predictions include cytoskeletal extracts with motors and ATP, suspensions of bacteria or pulsatile cells, or fluids laden with artificial self-propelled colloids.
\end{abstract}

\section{Introduction and results}
\label{intro}
\subsection{Background}
\label{backgrd}
The dynamics of suspensions of self-propelled particles is a subject of enduring interest in fluid mechanics \citep{pedleykessler,sanoop2006,cisneros,mehandia2008,ishikawareview,laugareview,ortizetal2009,ganeshannurev2011}. 
One line of recent progress in this field has come through the observation that ordered phases of active particles could be viewed as liquid crystals with a key novel feature: the constituent particles are endowed with self-generated ``active stresses'' \citep{tonertusr,tanniecristina2006,physrep,hfsp,aparnacristinaPNAS,srannurev,menonreview2010}. The resulting extension of liquid crystal hydrodynamics to include self-propelling stresses, initially proposed as a coarse-grained description of swimmers \citep{aditisr}, also emerged naturally in a theory of the cytoskeleton as an active gel \citep{activegel1}. Within liquid-crystal hydrodynamics one could imagine a wide variety of orientationally or translationally ordered states, but most recent work has focused on the case of local uniaxial orientational order. In such situations it is useful to distinguish active fluids as contractile or extensile, according to whether the active stress generates flow inward or outward along the local ordering axis. This classification corresponds to the puller-pusher classification at the scale of a single swimming particle \citep[see, e.g.,][]{saintillanexpmech}. However, for systems such as motor-filament extracts, where a pulling or pushing motif is not obvious at the scale of a single particle, the sign of the active stress at a coarse-grained scale is best established by an independent measurement \citep{takiguchi}. A further distinction is necessary, between polar states which break fore-aft symmetry along the axis of orientation, and apolar states which are fore-aft symmetric. Swimmers in a macroscopically coherent flock are in a state of polar order. However, to leading order in inverse powers of distance from the particle, i.e., to force-dipole order, the flow-field generated by a general swimming particle is the same as that produced by one that merely stirs its surroundings without propelling itself. The distinction between ``movers'' and ``shakers'' \citep{hatwalneetal2004} lies in their near field and is thus subdominant in a gradient expansion. When studying rheology \citep[see, e.g.,][]{hatwalneetal2004,tanniecristina2006,marenduzzo2008,aranson2009,rafai2010,saintillanexpmech} and large-scale flow, it is therefore a reasonable starting point to ignore the consequences of polar order in active fluids. Accordingly our study in this paper is restricted to states with \textit{apolar} uniaxial orientational order.  The effect of polarity on the instability of active fluids \citep{giomi}, the consequences of bottom-heaviness and more generally the interplay of swimming with gravity \citep{bioconvection} all lie outside the scope of this work. 

Instability and self-generated flow are conspicuous features of active fluids. Indeed, the idea of hydrodynamic instability in living matter, and even the term active stress, goes back at least to the work of \citet{finscriv}. Their primary interest, however, lay in cases in which the stress was built from tensorial combinations of gradients of local scalars such as concentration and temperature. The present work, and much recent work on flows in active matter, relies on the ordering of orientational degrees of freedom to transmit the effects of local activity to large scales. A bulk sample of active fluid or suspension in a state of uniaxial orientational order is under macroscopic uniaxial compression or tension, and must therefore buckle along or transverse to the axis. This is why the non-flowing state of a bulk active fluid, whether extensile or contractile, is in general unstable \citep{aditisr,voit} towards director distortion and flow without the assistance of externally imposed pressure gradients. 

Theoretical studies of active fluids in a thin-film geometry with one or more free surfaces predict particularly striking effects for a reference state in which the orientation field lies parallel to the film. \citet{voit2} consider a freely-suspended apolar film, ignore the displacements of the free surfaces, and predict the spontaneous generation of topological defect arrays. \citet{sumithra} consider an active film spread on a solid surface, and show that the coupling of \textit{polar} orientational order to the tilt of the free surface of the film leads to growing and propagating undulations. Both these studies consider films unbounded in the lateral plane. \textit{Finite} drops of suspensions of self-propelled organisms have been studied in theory and experiment by \citet{bees2000,bees2002}, but the key ingredients in those studies were cell division, the secretion of a wetting agent, and concentration dependence of the viscosity. 

In this paper we examine theoretically the effect of self-propelling stresses
on the shape and the spreading behaviour of a finite drop of active fluid
on a solid substrate, under conditions of partial wetting with small equilibrium contact angle, and local polarization anchored parallel to the surface of the drop. This boundary condition is in contrast to the perpendicular alignment frequently encountered in the liquid-crystal literature; see, e.g., \citet{gupta,yamaguchi,prost} 
\begin{figure}
\includegraphics[width=\columnwidth]{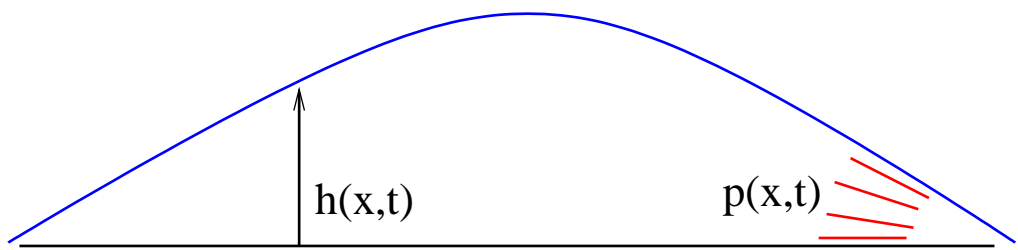}
\caption{A drop characterized by height and orientation fields $h$ and $\mathbf{p}$, the latter constrained to lie parallel to the bounding surfaces. For the present apolar case, $\mathbf{p}$ and $-\mathbf{p}$ are equivalent, so we depict the orientation by headless arrows. For clarity we have shown the filaments and the orientation field only near the edge of the drop.}
\label{dropfig}
\end{figure}
We consider separately the cases of contractile and extensile active stresses, and 
study two situations: steady standing drops, with vanishing velocity field, and 
spreading drops. 

\subsection{Summary of results}
\label{results}
Here are our main results in brief. (i) The shape of the standing drop is determined 
primarily by the interplay of active stresses $\sigma_0$ and surface tension $\gamma$, 
with the liquid-crystal elasticity of the ordered filaments playing a role near the contact line. 
The stationary height profile $h(x)$ as a function of the horizontal coordinate $x$ can be obtained as the Euler-Lagrange equation for position $h$, as a function of ``time'' $x$, of a particle in an effective potential $V(h) = {-\sigma_0 } (h \ln h / h_0 - h)$. 
(ii) For a 3-dimensional drop, the anchoring conditions impose the existence of defects in the polarization field. We propose three possible defect structures; both the final state of spreading and the spreading kinetics depend on the precise nature of these defects.  (iii) Extensile active stresses lead to flat drops, with thickness varying as  $\gamma \theta^2/\sigma_0$  where $\theta$ is the equilibrium contact angle. (iv) The interplay between active and viscous stresses leads to anomalously rapid spreading of the drop as a 
function of time $t$. In the simple case where the orientation field has an aster-like defect line at the center of the drop, the drop retains rotational symmetry with diameter $R \sim (\sigma_0 \Omega^2 t /\mu)^{1/6}$ 
where $\Omega$ and $\mu$ are its volume and shear viscosity respectively. For a $1+1$-dimensional geometry, we predict a $t^{1/4}$ growth law. (v) For a vortex defect, in which the orientation axis points in the azimuthal direction, spreading is arrested at long enough times, and the final height profile of the drop is a non-monotone function of radius as shown in Eq. (\ref{vortex shape}) and Fig. \ref{figvortex}. For contractile active stresses the drop is fatter than in the absence of activity (Fig. \ref{contractile}). 

\begin{figure}
\includegraphics[width=\columnwidth]{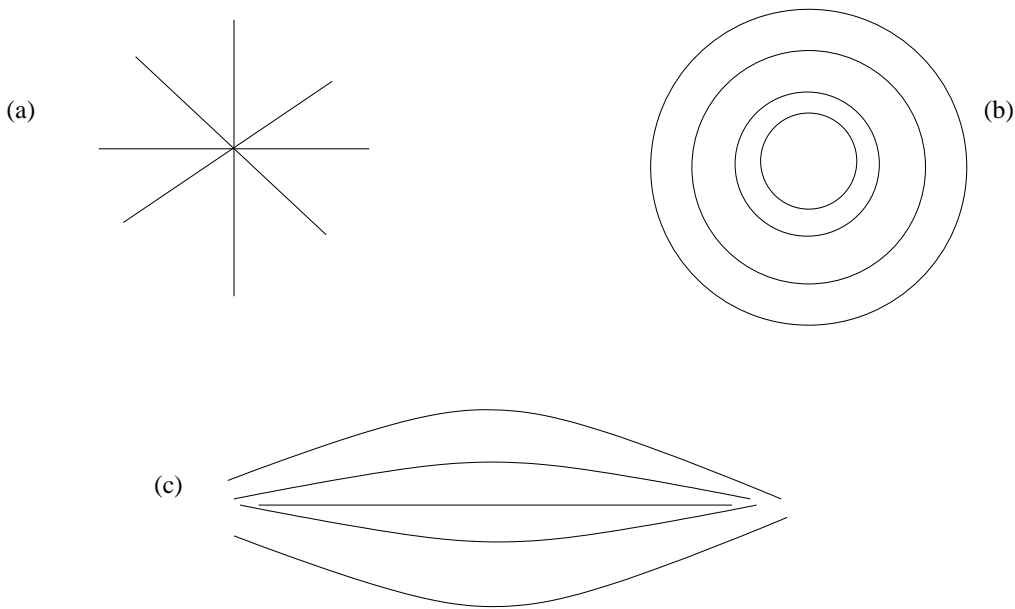}
\caption{Sketches of the three defect configurations of the nematic director field that we consider: (a) the aster, (b) the vortex and (c) the pair of half-defects.}
\label{3defects}
\end{figure}
\section{Detailed theory}
\label{calcdetails}
We now describe our calculations in more detail. Consider as in Fig. \ref{dropfig} 
a drop with a shape given by its thickness $h(\rp,t)$ in the vertical ($z$) direction  as a function of in-plane position $\rp = (x,y)$ at time $t$ and containing filaments with orientation field $\mathbf{p}$. The active character of the fluid endows the drop with an active contribution  \citep{aditisr,activegel1} to the stress tensor of the medium.
\begin{equation}
\label{sigact}
 \sigma^a = 
- \sigma_0 \mathbf{p}\mathbf{p}
\end{equation}
Negative and positive values of $\sigma_0$ correspond respectively to contractile and extensile active stresses, and the magnitude of $\sigma_0$ is proportional to the local number density of filaments $c$, which for simplicity we assume to be constant and uniform, although interesting physical effects are likely to arise if the dynamics of $c$ is taken into account. The anchoring condition is such that there is  no normal component of the polarization at any bounding surface with which the suspension is in contact. This boundary condition is motivated in part by observations \citep{kemkemer2000} on thin films of amoeboid cells, in which the cells lie in the plane of the glass slide on which they are spread, and form nematic liquid crystalline structures. Although we use a vectorial orientation field, in this paper we only discuss apolar fluids where $\mathbf{p}$ and $-\mathbf{p}$ are equivalent. This arises if the filaments lack a head-tail distinction, or are ordered with axes parallel on average but without respecting the head-tail polarity. Our aim is to obtain equations of motion for $h$ and a $z$-averaged orientation $\mathbf{p}(\rp,t)$, whose static solutions yield the steady-state shape of the drop and whose dynamic scaling properties lead to the spreading laws for the drop \citep{leger,degennes}. Our treatment is along the lines of \citet{benamar}, taking into account, as in \citet{sumithra}, the presence of active stresses but ignoring the effects of polarity, and applied to a drop rather than an unbounded film. 

We solve for the hydrodynamic velocity field $\mathbf{u}= (\mathbf{u}_\perp, u_z)$ 
inside the drop for a given thickness profile, and obtain the equations of motion for $h$ 
from the kinematic boundary condition \citep{stone} $ \dot h = 
u_z - \mathbf{u_\perp}\cdot \mathbf{ \nabla_\perp} h$ connecting the thickness $h$ 
to the velocity field $\mathbf{u}$ evaluated at the free surface. The condition of incompressibility $\mathbf{\nabla} \cdot \mathbf{u}$ = 0 turns the height dynamics into a local conservation law 
\begin{equation}
\label{hincompeqn}
\partial_t h + \mathbf{\nabla}_\perp \cdot \int_0^h \mathbf{u}_\perp dz 
= 0
\end{equation}
in the $\perp$ plane. 
For simplicity, we only consider here the case where the equilibrium contact angle of the drop $\theta$ is small. 
As appropriate for slow, small-scale thin-film flows we obtain the thickness-averaged horizontal velocity in (\ref{hincompeqn}) in terms of $\mathbf{p}$, $c$ and $h$ by solving the steady Stokes equation which, in the lubrication approximation \citep{stone} $u_z=0$, $|\nabla_{\perp} \mathbf{u}|\ll |\partial_z \mathbf{u}|$, reads 
\begin{equation}
\label{fluideom}
\mu \partial_z^2 \mathbf{u}_\perp -
\mathbf{\nabla} P
+ \mathbf{\nabla} \cdot (\mathbf{\sigma^a} + \mathbf{\sigma^e}) = 0 
\end{equation}
where $\mu$ and $P$ are the viscosity and pressure field of the fluid, 
$\mathbf{\sigma}^e$ is the contribution to the stress from the nematic elasticity 
of $\mathbf{p}$  \citep[see, e.g.,][]{leslie1979,chandrasekhar1992,prost} and $\sigma^a$ is the active 
stress introduced above.

\subsection{1+1-dimensional treatment}
\label{1+1}
It is instructive to solve the problem in a simple $(1+1)$-dimensional picture of $h$ as a function of horizontal position $x$ and time $t$. We do this first, and later extend our discussion to determine shape and spreading kinetics in the $(2+1)$-dimensional case. As illustrated in Fig. \ref{dropfig}, the $z$ dependence of $p_z$ is determined by Frank elasticity \citep{prost} with boundary conditions $p_z =0$ at $z=0$ and $p_z = \partial_xh \ll 1$ at the free surface $z=h$, so that \citep{sumithra} 
\beq
\label{pzlininterp}
p_z = (z/h)\partial_x h. 
\eeq
The $z$ component of the equation of motion 
(\ref{fluideom}) reads $-\sigma_0\partial_x p_z = \partial_zP$ 
where $P$ is the pressure. Expressing $p_z$ in terms of $h$ via (\ref{pzlininterp}), 
using the Laplace pressure boundary condition with surface tension $\gamma$, 
and solving for $P$ we find \citep{sumithra} 
\beq
\label{pressureinfilm}
P = P_0 - \gamma  \partial_x^2h + P_a (z) 
\eeq
with  
$P_a(z)= -(\sigma_0/2)(z^2-h^2)
(h''/h - h'^2/h^2)$ where a prime ($'$) denotes $\partial_x$.
In the following we ignore this active contribution to the pressure, which within the lubrication approximation gives only a negligible contribution to the equation of motion below.

%
The equation of motion is obtained by integration of the $x$ 
component of Eq. (\ref{fluideom}), using (\ref{pressureinfilm}),  the active stress from (\ref{sigact})  and  including in the stress the contribution of nematic 
elasticity with the orientation profile (\ref{pzlininterp}) and a single Frank constant 
$K$.
\beq
\label{heighteom}
\mu  \bar{\mathbf{u}}_\perp=\frac{h^2}{3}\left( \gamma  \partial_x^3h -\partial_x \frac{\delta F_{el}}{\delta h}- \frac{\sigma_0}{h} \partial_x h \right)
\eeq
where $\bar{\mathbf{u}}_\perp $ is the inplane velocity averaged over the 
thickness of the film and 
$F_{el}= (K/2)\int dx ( \partial_x h)^2/h$
is the nematic elastic free energy. Combining Eq \ref{heighteom}, with the mass 
conservation Eq. \ref{hincompeqn}, one obtains an independent equation
for the height field $h$. 
\beq
\label{isoheighteom}
\partial_t h + \frac{1}{ 3 \mu} \nabla\cdot (\gamma  h^3 \nabla \nabla^2 h-
\nabla h^3 \nabla \frac{\delta F_{el}}{\delta h}
- \sigma_0  h^2 \nabla h) = 0
\eeq
where the gradient is taken here is the $x$ direction.

The effects of activity on spreading enter at the same order in gradients as those 
of gravity, but with a different dependence on height, essentially because of the existence 
of an intrinsic stress scale $\sigma_0$. Indeed, the contribution of activity to 
(\ref{heighteom}) defines an active velocity  $v_{act} =  -M_h \partial_x P_{act}$, 
with the hydrodynamic mobility $M_h = h^2/(3\eta)$ and an active disjoining pressure 
\beq
\label{activepressure}
P_{act} = \sigma_0 \ln h/h_0
\eeq
 where $h_0$ is an arbitrary reference height. 

\begin{figure}
  \begin{center}
	\includegraphics[width=\columnwidth]{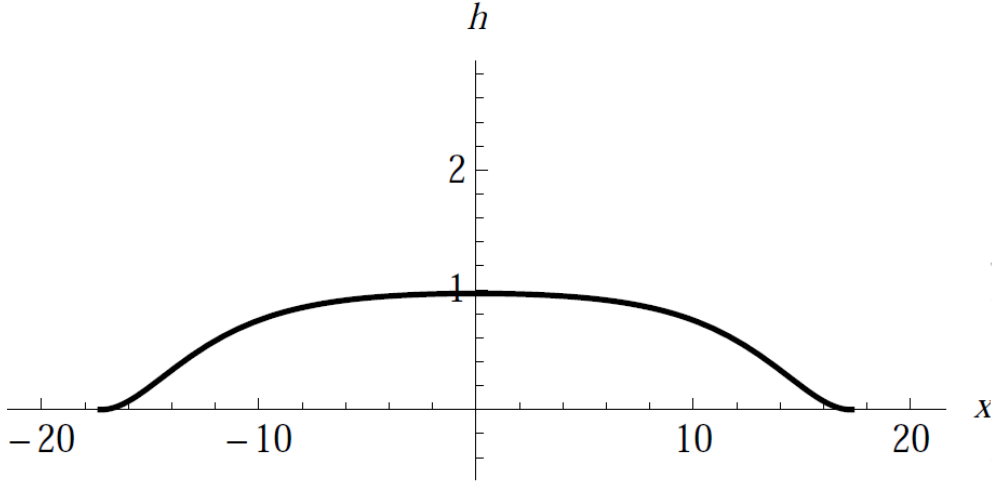}	
	\caption{\label{elephantfig} Height profile of a spread extensile drop, flat on top with a precursor region of quadratic variation at the limbs, obtained from a numerical solution of (\ref{static1d}) for drop of finite volume fixed by a pressure of .0001 in units of $\sigma_0$.}
  \end{center}
\end{figure}

\begin{figure}
  \begin{center}
	\includegraphics[width=\columnwidth]{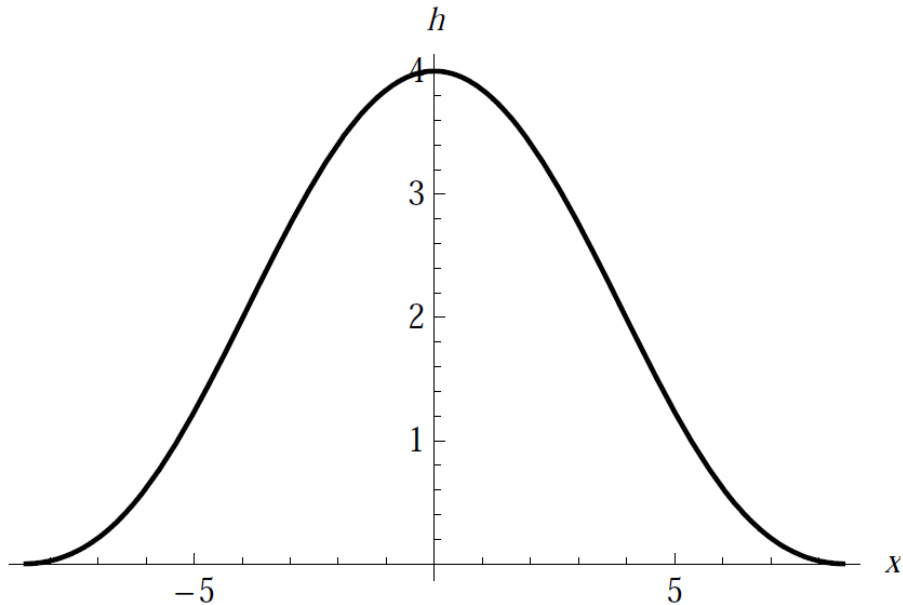}	
	\caption{\label{contractile} Height profile of a spread contractile drop, higher and with a smaller base than in the extensile case (Fig. \ref{elephantfig}), obtained from a numerical solution of (\ref{static1d}) with the sign of the active stress reversed.}
  \end{center}
\end{figure}

The shape of the static drop configuration that results at the end of spreading is particularly 
easy to understand in the $1+1$-dimensional picture. In the static limit (\ref{heighteom}) integrated once with respect to $x$ yields 
\beq
\label{static1d}
{\gamma}\partial^2_x h -{\sigma_0 } \ln {h \over h_0} - \frac{\delta F_{el}}{\delta h}= \mbox{const.} = 0
\eeq
where $h_0$ is the asymptotic maximum height for an infinite drop. The integration 
constant on the right hand side, i.e., the negative of the pressure jump at the top of the drop, has been taken to be zero, assuming a macroscopic flat drop. For a finite size drop, this pressure is obtained from the volume conservation constraint. Multiplying by $\partial_x h$ and integrating once again with respect to $x$ yields, 
after some algebra, 
\beq
\label{pseudoenergy}
\frac{1}{2}({K \over h} + \gamma)( \partial_x h)^2 -\sigma_0 (h \ln {h \over h_0} - h) 
= \mbox{const.} = {\gamma \over 2} \theta^2
\eeq
where $\theta$ is the Young contact angle that the drop would achieve in the absence
of activity. It can be seen as the effective contact angle made by the drop at the solid 
surface, for heights approaching the value $h_K \sim K/\gamma$ below which 
nematic elasticity becomes important. Above this height we can neglect $K/h$, and (\ref{pseudoenergy}) takes the form of energy conservation for a particle of mass $\gamma$, position $h(x)$ as a function of time $x$, moving in a potential 
\beq
\label{veff}
V(h) = {-\sigma_0 } (h \ln{h \over h_0} - h). 
\eeq
which vanishes at $0$ and has a maximum at $h_0$. The steep growth of 
$V(h)$ from $h=0$ implies a very slow progress of the effective particle, which means 
the height changes very slowly with $x$ for $h > h_K$, i.e., a flat free surface.  
The maximum height corresponds to the classical turning point for motion in the 
potential $V(h)$ (\ref{veff}), $\partial_x h=0$ at $h=h_0$. From (\ref{pseudoenergy}) 
this means 
\beq
\label{hzero}
h_0 = {\gamma \over 2} {\theta^2 \over \sigma_0}. 
\eeq
The drop is flat with a thickness $h_0$ over most of its area. At its edge, the thickness 
decays 
over a thickness $\ell \sim h_0/\theta \sim \gamma \theta/\sigma_0$. The condition 
for the drop to be flat when it reaches equilibrium is that its final radius $R$ be much 
larger than the size $\ell$ of the edge.
For $h \to 0$ the shape of the spreading drop is driven by the Frank elasticity of the polarization 
in (\ref{pseudoenergy}): 
\beq
\label{hto0}
\frac{K}{2h}(\partial_x h)^2 \simeq {\gamma \over 2} \theta^2
\eeq
so that 
$h \sim (\gamma \theta^2 /4 K)x^2$,
growing quadratically from zero near the left 
edge which we have take n to be at $0$, and similarly for the right edge. The resulting 
height profile is sketched in Fig. \ref{elephantfig}.

The profile calculated and presented in (\ref{veff})-(\ref{hto0}) and Fig. \ref{elephantfig}
is consistent only if the height of the drop $h_0$ is larger than the thickness $h_K = K/\gamma$ below which nematic elasticity becomes relevant.  This holds true if $\theta^2> \sigma_0 K/\gamma^2$. If this condition is not satisfied the wetting properties of the drop are not dominated by the active stress but by the nematic elasticity. This regime has been studied both experimentally and theoretically by \citet{cazabat}, and we do not consider it any further. As mentioned in the introduction, a thin film of active fluid becomes unstable via a Freedericksz-like instability and flows spontaneously \citep{voit} if its thickness is larger than a critical thickness $h_c \sim (K/\sigma_0)^{1/2}$. A non-flowing steady state of a spread active drop can therefore exist only if its height $h_0<h_c$. To observe the effects we are discussing we must then have $h_K<h_c$, i.e., $\theta^4<\sigma_0 K/\gamma^2$. Note that this condition is compatible with  the condition that the wetting is dominated by the active stresses only if $\sigma_0 K/\gamma^2 \ll 1$. We assume that this condition is satisfied in the following so that that the thin film formed by the active drop does not show any Freedericksz instability.

For a contractile drop, $\sigma_0 < 0$, the situation is less illuminating. For $h < h_K$ 
the height profile is as in the extensile case, viz., quadratic growth of the height, as the active 
stress $\sigma_0$ plays no role and the Frank energy dominates. Thereafter, we repeat 
the analysis of motion in the effective potential $V(h)$ (\ref{veff}), which 
now \textit{decreases} with an infinite slope at $h=0$, passes through a minimum at 
$h_0$ and increases thereafter. The height profile is thus higher and fatter than that 
of a sessile drop in the absence of activity.

We now consider the spreading kinetics of a $1+1$-dimensional active drop in its late stages where its thickness is smaller than the critical thickness $h_c$ for the appearance of spontaneous flow. The equation of motion for a spreading drop 
is given by Eq.\ref{isoheighteom} where we neglect the nematic elasticity term that only plays a role at very small thicknesses.
This spreading equation can be discussed at the scaling level. Except for the very edge of the 
drop, the surface tension term is small compared to the active stress term. This leads to
${dh_0}/{dt} \sim -\sigma_0 {h_0^3}/{R^2}$ where $h_0$ is the maximum
thickness at the center of the drop and $R$ its radius. The two-dimensional volume of the 
drop $\Omega \sim h_0 R$ is conserved. We therefore obtain the spreading law
\begin{equation}
\label{1d}
 R(t) \sim \left(\frac{\sigma_0 \Omega^2 t}{\mu}\right)^{1/4}
\end{equation}

\subsection{Three-dimensional treatment: topological defects}
\label{3d}
We now consider the real three dimensional geometry and discuss  the spreading kinetics 
of an active drop. The anchoring conditions that we impose can be satisfied only if there 
exist defects in the nematic polarization field \citep{prost}. We consider three types of defects: 
an aster-like cylindrical wedge disclination located at the center of the drop with polarization pointing radially from the defect line; a vortex-like wedge disclination located at 
the centre of the drop with polarization aligned in concentric circles in the azimuthal direction around the defect line; and two point-like half-defects located on the substrate on opposite 
sides of the drop as shown in Fig. \ref{3defects}. The first two topologies are 
simpler because the active drop remains axially symmetric but it is now well established 
that aster-like line defects are unstable and degenerate into two point-like half-defects. If the 
two defects remain on the vertical axis, the deformation of the polarization is virtually 
the same at that obtained for the original line defect. It is more likely that the 
two point-like defects migrate to opposite sides of the drop on the substrate which is 
our last topology.

For the aster-like line defect (Fig. \ref{3defects}a), the polarization is mostly  radial with a small component 
along the $z$ direction. This is very similar to the $1+1$ geometry studied above and if 
we ignore the core of the defect, 
the derivation of the equations for the local thickness of the drop exactly follows 
the same lines. The final shape of the drop is a circular pancake with a thickness given 
by equation \ref{hzero}. The equation governing the thickness during spreading is still Eq.
\ref{isoheighteom} and a scaling analysis imposing a constant volume of the drop 
$\Omega \sim hR^2$ leads to the spreading law:
\begin{equation}
 \label{2d}
 R(t) \sim \left(\frac{\sigma_0 \Omega^2 t}{\mu}\right)^{1/6}
\end{equation}

For a vortex-like line-defect (Fig. \ref{3defects}b), the polarization is always strictly azimuthal. In polar coordinates 
($r, \theta$), it is always along the $\theta$ direction and the only non vanishing 
component of the active stress is $\sigma_a^{\theta \theta}=-\sigma_0$. The equation 
of motion in this case written in polar coordinates is
\beq
\label{vortex}
\mu  \bar{\mathbf{u}}_\perp=\frac{h^2}{3}( \gamma \nabla \nabla^2 h +
\frac{\sigma_0}{r} {\hat r}) = 0
\eeq
where ${\hat r}$ is the unit vector along $r$.
The active stress appears here as similar to a centrifugal force with an unusual 
$r$ dependence $\sim 1/r$ instead of $\sim r$.
\begin{center}
\begin{figure}
\includegraphics[width=\columnwidth]{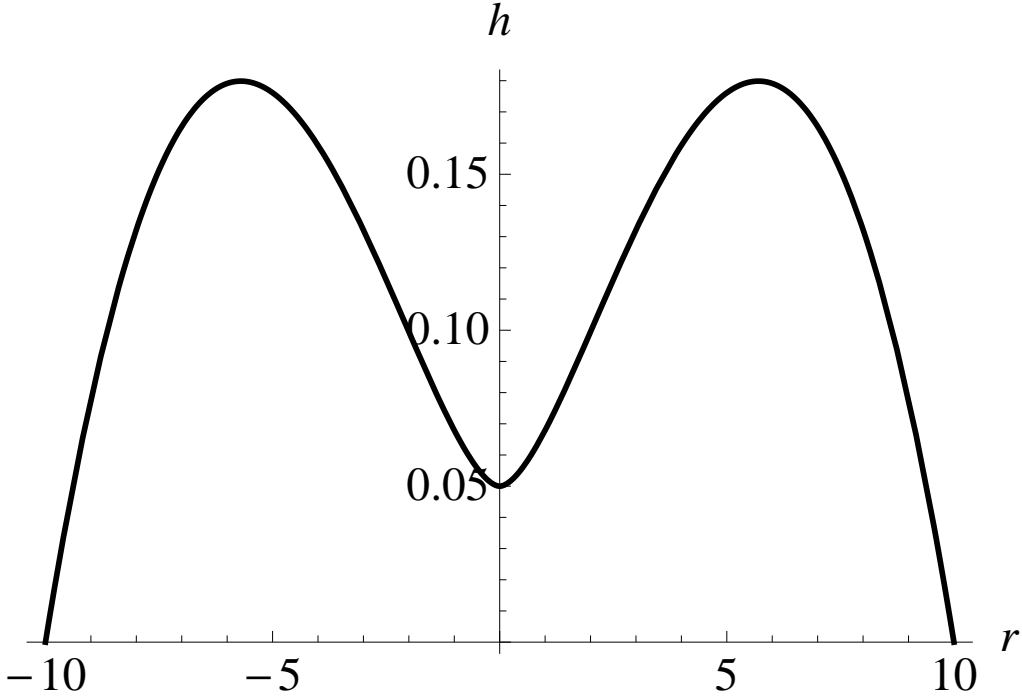}
\caption{The final shape, from Eq. (\ref{vortex shape}), of an active drop in which the director field is in the vortex configuration, oriented strictly in the azimuthal direction, with a singularity on a vertical line segment at the centre of the drop.}
\label{figvortex}
\end{figure}
\end{center}
The final stationary shape of the drop can be calculated explicitly by imposing a
vanishing velocity on (\ref{vortex}): 
\begin{equation}
 \label{vortex shape}
 h= -\frac{\sigma_0 R^2}{4\gamma} r^2 \log r/R -\frac{\sigma_0}{8\gamma}(R^2-r^2) 
 + \frac{\theta}{2R} (R^2-r^2)
\end{equation}
where $R$ is the radius of the drop. This profile is shown in Fig. \ref{figvortex}. The thickness has a minimum at the center of the drop where the polarization gradient is large. It has a maximum at an intermediate radius $r=R \exp 
-(2\gamma\theta /\sigma_0 R)$. There does not seem to exist any simple scaling regime for the spreading kinetics of the drop in this case except at early times when the the active effects are negligible and the drop spreads as a simple liquid drop $R(t) \sim t^{1/10}$. The study of the spreading kinetics would require a numerical study of the spreading equations that we postpone for future work.

The third topology with opposite defects on the two sides of the drop, as in Fig. \ref{3defects}(c),  is more complex since in this case the drop is anisotropic and spreads preferentially in the direction of the two defects and its radius in this direction $R_x$ is larger than its radius in the 
perpendicular direction along the planar substrate $R_y$. As above, we suppose here
that the polarization field follows adiabatically the deformation of the drop and reaches rapidly a local equilibrium where the molecular field conjugate to the polarization vanishes. 
When the anisotropy $R_x/R_y$ is large, the polarization can be estimated from nematic elasticity as 
$p_x\simeq 1$, $p_y \simeq -xy/R^2$ and $p_z =
(z/h) \partial_x h$ over most of the volume of the drop. Inserting 
these results in the expression of the active 
stress one derives a dynamic equation equivalent to (\ref{isoheighteom}). Estimating 
the order of magnitude each term in this dynamic equation, 
we obtain two scaling relations   $(d R_x / dt)/R_x \sim (d R_y / dt)/R_y \sim \sigma_0 h^2 / R^2$. Together with volume conservation $hR^2=$ constant, these equations do not suffice to obtain a growth law for the spreading drop as they do not give information on the anisotropy: this would require the knowledge of the prefactors of the scaling relations. 
Two possible limiting cases are (i) finite, bounded anisotropy, where the active drop is characterized by a single length-scale given by (\ref{2d}) and (ii) infinite anisotropy, where the drop can be considered as one dimensional with a spreading law given by (\ref{1d}). Further progress requires a numerical solution, which we do not attempt here.

\section{Summary and outlook}
\label{summary}
In closing, we summarise our approach and results, and offer a perspective on future directions. From the point of view of fluid mechanics, the distinguishing feature of suspensions of self-driven particles is the permanent uniaxial stress with which each particle is endowed. Long-range orientational order organises the individual active stresses, transmitting their effect to macroscopic scales and leading to consequences for large-scale fluid flow. The spreading of a liquid drop on a solid surface, driven by gradients in the disjoining pressure, is a fundamental problem in fluid mechanics. Active stresses provide a hitherto unexplored contribution to disjoining pressure. In an active suspension with nematic liquid-crystalline order, the deviatoric active stress is determined by the local nematic director which, in our treatment, is linked to the shape of the drop through the requirement that there is no normal component of the orientation field at any bounding surface with which the suspension is in contact. Thus at the base of the drop the director is purely horizontal, while variations in the free-surface height tilt the orientation. The resulting inhomogeneities in the active stress generate flows that drive the spreading process with novel growth laws and drop shapes. The closed geometry of a drop constrains the director field, inducing topological defects. The interplay of spreading and defects has not been studied even in conventional equilibrium nematics, as far as we know. We have studied $1+1$ and $2+1$ dimensional geometries. In a $1+1$-dimensional geometry, corresponding to a distended drop, we are able to predict an anomalously flat final shape and a $t^{1/4}$ spreading law for extensile active stresses. For a full $2+1$-dimensional case, scaling alone offers only partial information for the case where a pair of half-defects are situated at the ends of the drop. For the case of a single defect line at centre, we find a $t^{1/6}$ spreading law for a radial ``aster'', and arrested spreading with a height profile with a non-monotone dependence for a vortex configuration, in which the director is strictly azimuthal. The complete $2+1$-dimensional spreading law requires a numerical solution of the active thin-film equations. 

Before we end, let us note some of the simplifying assumptions we have made, some of which need to be relaxed in a more complete treatment. First is the approximation of a small contact angle. For our boundary condition, namely, filaments parallel to the bounding surface with which they are in contact, the effects of activity on spreading are strongest in the limit of a small contact angle. A contact angle close to 90$^{\circ}$ would mean that filaments lying along the free surface but close to the base would push vertically rather than laterally, suppressing the contribution of active stresses. Elastic anisotropy, which we have ignored, should lead to spiral defect configurations which should rotate spontaneously \citep{activegel1}. If we take into account the polar nature of the order parameter, even a pure vortex should rotate, and a macroscopically polar active drop should translate. We have not examined so far the stability of active drops with respect to azimuthally varying perturbations. Finally, \textit{in vivo} realizations of active drops in general have complicating features that will need to be suppressed experimentally or, eventually, included in a comprehensive theoretical analysis. For example, cell division will modify spreading in bacterial suspensions \citep{bees2000,bees2002}. From a long-term perspective, our work is of relevance to an understanding of the spreading of a single cell. However, in that setting, treadmilling \citep{treadmilling1999}, that is, the polymerization and depolymerization of actin filaments, will play an important role. Notwithstanding these limitations, we look forward to experimental tests of our predictions, on microorganism suspensions, extracts, or artificial self-propelled systems \citep[see, e.g.,][]{paxton2004,howse2007,bartolo2010}. 

We close by expressing our pleasure at being able to contribute to a field which Tim Pedley has defined, led, and illuminated. 

JF and SR were supported by grant 3504-2 of the Indo-French Centre for the Promotion of Advanced Research, and SR by the Department of Science and Technology, India through a J C Bose Fellowship and the Centre for Mathematical Biology (Grant No. SR/S4/MS:419/07). 
\bibliographystyle{jfm}
\bibliography{activedropjfm_final}
\end{document}